\documentclass[12pt,a4paper]{article}
\usepackage{epsfig}
\def\eq#1{{Eq.~(\ref{#1})}}

\newcommand{\beq}{\begin{equation}}
\newcommand{\eeq}{\end{equation}}
\newcommand{\beqar}[1]{\begin{eqnarray}\label{#1}}
\newcommand{\eeqar}{\end{eqnarray}}
\def\arnps#1#2#3{  {\it Ann. Rev. Nucl. Part. Sci. }{\bf #1}:#2 (#3)}
\def\npb#1#2#3{    {\it Nucl. Phys. }{\bf B#1}:#2 (#3)}
\def\plb#1#2#3{    {\it Phys. Lett. }{\bf B#1}:#2 (#3)}
\def\prd#1#2#3{    {\it Phys. Rev. }{\bf D#1}:#2 (#3)}
\def\prep#1#2#3{   {\it Phys. Rep. }{\bf #1}:#2 (#3)}

\def\zpc#1#2#3{    {\it Z. Phys. }{\bf C#1}:#2 (#3)}

\def\sjnp#1#2#3{   {\it Sov. J. Nucl. Phys. }{\bf #1}:#2 (#3)}

\def\jetpl#1#2#3{  {\it JETP Lett. }{\bf #1}:#2 (#3)}

\begin{document}
\title {{\bf Higher twist corrections  and maxima for DIS }\\
{\bf   on a proton in the high density QCD region}}
\author{
{\bf
E.~Gotsman,\thanks{e-mail: gotsman@post.tau.ac.il}~$\, ^a$,
\quad  E.~Levin,\thanks{e-mail: leving@post.tau.ac.il}~$\, ^a$,
\quad U.Maor,\thanks{e-mail: maor@post.tau.ac.il}~$\, ^a$,
} \\
{\bf
L.~McLerran,\thanks{e-mail: mclerran@bnl.gov}~$\, ^b$,
 \quad
K.~Tuchin\thanks{e-mail: tuchin@post.tau.ac.il}~$\, ^a$,
} \\[10mm]
{\it\normalsize $^a$HEP Department}\\
 {\it\normalsize School of Physics and Astronomy,}\\
 {\it\normalsize Raymond and Beverly Sackler Faculty of Exact Science,}\\
 {\it\normalsize Tel-Aviv University, Ramat Aviv, 69978, Israel}\\[0.5cm]
{\it\normalsize $^b$ Physics Department,}\\
{\it\normalsize Brookhaven National Laboratory,}\\
 {\it\normalsize Upton, NY 11973-5000, USA}
}

\date{}
\maketitle
\thispagestyle{empty}

\begin{abstract}
We show that the ratio of different structure functions have a
maximum which depends on energy. We argue, using the Golec-Biernat and
Wusthoff model as well as the eikonal approach, that these maxima
are functions of the saturation scale. We analyze leading and
higher twist contributions for different observables to check
whether a kinematic region exists where high parton density
effects can be detected experimentally.

\end{abstract}
\thispagestyle{empty}
\begin{flushright}
\vspace{-17.5cm}
TAUP-2644-2000\\
BNL-NT-00/20\\
\today
\end{flushright}

\newpage
\setcounter{page}{1}



\section{Introduction}
In this letter we continue our study of phenomena associated with
gluon saturation in deep inelastic scattering (DIS) following our
original study
in Ref.~\cite{WE}. Our goal is to find the kinematic region where
the evolution of parton densities can no longer be
 correctly  described by
linear evolution equations\cite{GLR,MUQI,MCVE,HDPQCD}.
 To solve this problem analytically one
has to find a solution of the non-linear evolution equation which
was derived in Refs.~\cite{KOV,BALITSKY}\footnote{Strictly speaking this
equation is derived for a nucleus target, however there are
reasons to assume\cite{BRAZ1} that it holds for the nucleon as
well.} and in Ref.~\cite{CLASSIC} in different approaches.
 In spite of the fact that the asymptotic solutions are
known\cite{LT}, a detailed description of the boundary of the
saturation region is still lacking. We would  like to determine how
gluon saturation can be detected in current and future
experimental data.

The idea suggested in Ref.~\cite{WE} was that there is a breakdown
of the Operator Product Expansion as one gets sufficiently close
to the saturation region. 
This enabled us to find the scale $\hat Q^2$
where the twist expansion breaks down. Since, at sufficiently
large $Q^2$, the saturation scale is the only scale of the
problem, it should be expressed unambiguously  in terms of $\hat Q^2$.
We refer the reader to our previous publication for more details.

In Ref.~\cite{WE} we used the eikonal
model\cite{DIPOLE,MU94,NN,DIX,Nucleon} to calculate the various
structure functions relevant in DIS on nuclei, and then to expand
them into the twist series. In this letter we will address the
subject of DIS on a nucleon and  apply the same approach developed
for DIS on nuclei.

The paper is organized as follows: We start with a brief review of
our approach without going into details of the calculation (Sec.~2). In
Sec.~3, we discuss numerical estimates regarding the nucleon
target. Finally, in Sec.~4 we conclude by proposing possible ways
to measure the different twists at HERA and LHC.

\section{The Model}

As was pointed out in the introduction we use the eikonal (or
Mueller-Glauber) model to calculate shadowing corrections. A
complete description of the eikonal  approach to shadowing
corrections (SC) was given in Ref.~\cite{Nucleon}. It was shown in
Ref.~\cite{Nucleon,NUCLEUS, BRAZ} that the eikonal approximation
leads to a  reasonable theoretical as well as experimental
approximation in the region of not too small $x_B$. There are
weighty arguments in favor of using the eikonal model to calculate
contributions of different twists\cite{WE} in various cross
sections. This simple  eikonal model is a reasonable starting
point for understanding how the shadowing corrections occur.
 The use of the Glauber-Mueller approach was justified for a
nuclear target\cite{KOV}, and it should hold for a nucleon as
well\cite{BRAZ1}.

Following Ref.~\cite{WE} and assuming that all dipoles in the
virtual photon wave function have the size $r_t^2=4/Q^2$ and
neglecting the quark masses  we obtain for the longitudinal and
transverse photon cross sections
\begin{eqnarray}
\sigma_L(\gamma^*p)
&=& 2\alpha_{em}\pi R^2\frac{Q_s^2}{4Q^2} G^{3\, 2}_{3\, 4}
\left(\frac{Q^2}{Q_s^2}\Bigg|{0,1;\frac{5}{2}\atop 1,1,2;-\frac{1}{2}}\right)
\label{LONG}\\
\sigma_T(\gamma^*p)&=&4\alpha_{em}\pi R_A^2\frac{Q_s^2}{4Q^2}
\left[ G^{3\, 2}_{3\, 4}
\left(\frac{Q^2}{Q_s^2}\Bigg|{1,1;\frac{5}{2}\atop 1,1,1;\frac{1}{2}}\right)
-\frac{1}{2}
G^{3\, 2}_{3\, 4}
\left(\frac{Q^2}{Q_s^2}\Bigg|{0,1;\frac{5}{2}\atop 1,1,1;-\frac{1}{2}}\right)
\right.
\nonumber\\[3mm]
&&\left. +\frac{Q_s^2}{Q^2}\left\{\frac{1}{4}
G^{3\, 2}_{3\, 4}
\left(\frac{Q^2}{Q_s^2}\Bigg|{1,1;\frac{7}{2}\atop 2,2,2;\frac{1}{2}}\right)
-\frac{1}{2}
G^{3\, 2}_{3\, 4}
\left(\frac{Q^2}{Q_s^2}\Bigg|{1,2;\frac{7}{2}\atop 2,2,2;\frac{3}{2}}\right)
\right\}\right]\label{TRANS}
\end{eqnarray}
where ${_3}F_1$ is the generalized hypergeometric function,
$G^{3\, 2}_{3\, 4}$ is the  Meijer function (see
Ref.\cite{BATEMAN}), and we have defined\footnote{In
our previous work \cite{WE} we used different notation $4 C=Q_s^2$.}
\beq\label{C}
Q_s^2(x_B,Q^2)\equiv\frac{2\pi\alpha_S\, x_BG(x_B,Q^2)}{3R^2}\quad
,
\eeq
where $G(x_B,Q^2)$ denotes the gluon distribution in the
nucleon,  $R$ the radius of the nucleon, and $Q_s^2$ is used as a
shorthand for $Q_s^2(x_B,Q^2)$ defined in \eq{C}.  $Q_s^2$ is a
saturation scale in the framework  of our model.

Diffractive cross sections can be calculated using the simple relations
following from the unitarity constraint:
\beq\label{DIFFRACT}
\sigma(\gamma^*p)^D_{T,L}(Q_s^2/Q^2)=
\sigma(\gamma^*p)_{T,L}(Q_s^2/Q^2)
-\frac{1}{2}\sigma(\gamma^*p)_{T,L}(2Q_s^2/Q^2)
\eeq

Formulae \eq{LONG}, \eq{TRANS} and \eq{DIFFRACT} make it possible
to estimate the contribution of any order twist to the particular cross
section\cite{WE} by closing the contour of integration in a Meijer
function over the corresponding pole.

We argued in the Ref.~\cite{WE}, that studying ratios of the cross
sections can yield very useful information about the saturation
region. While the cross sections receive large corrections from
the next-to-leading order of perturbative expansion, their ratios
do not\cite{Nucleon,BRAZ}. In this paper we will consider two
ratios $\sigma_L/\sigma_T$ and $\sigma_L^D/\sigma_T^D$. Both of
these ratios display a remarkable property as a function of photon
virtuality $Q^2$. At small $Q^2$ the ratios vanish since the real
photon is transverse. At large $Q^2$ they vanish as well, as
implied by pQCD. This leads us to suggest that both ratios have
maximum $Q_{max}^2$. We would expect that this maximum is a
certain function of the saturation scale, since at small
$x_B$ and large $Q^2$ this is the only scale important in the
process of dipole scattering. So, studying the behaviour of the
$Q_{max}^2$
with energy we can determine the $Q_s^2(x_B)$ as well. Two
problems occur. First, the confinement scale affects the position
of the maximum of ratios. We can only hope that at large energies
(small $x_B$) the maximum will occur at sufficiently large $Q^2$
where this influence is small, though as we will see below, it
cannot be completely excluded.

Second, we do not know the
precise form of the function $Q_s^2=f(Q_{max}^2)$ which has to be
determined using the full twist expansion. We can, however, roughly
estimate the form of this function using a simple approach.
Namely, we illustrate the most important features (for finding the
$Q_{max}^2$) of \eq{LONG} and \eq{TRANS} by the
following formulae which have the correct asymptotic
behaviour for $Q^2\ll m_f^2<Q_s^2$ and $Q^2\gg Q_s^2>m_f^2$, and which
are valid for any model:
\begin{eqnarray*}
\sigma_T &\sim& R^2\frac{Q_s^2}{Q^2+Q_s^2}\quad,\\
\sigma_L &\sim&
R^2\frac{Q^2}{Q^2+m^2_f}\cdot\frac{Q_s^2}{Q^2+Q_s^2}\quad,\\
\sigma_T^D &\sim& R^2\frac{Q_s^2}{Q^2+Q_s^2}\quad,\\
\sigma_L^D &\sim&
R^2\frac{Q^2}{Q^2+m^2_f}\cdot\frac{Q_s^4}{(Q^2+Q_s^2)^2}\quad,
\end{eqnarray*}
where the logarithmic contributions are neglected.  Note, that
we explicitly assumed here that at large energies the only
relevant scale of the process is $Q_s^2$.

The ratios of the cross sections  then read
\begin{eqnarray*}
R &=&\frac{\sigma_L}{\sigma_T}=\frac{Q^2}{Q^2+m_f^2}\quad ,\\
R^D &=& \frac{Q^2}{Q^2+m_f^2}\cdot\frac{Q_s^2}{Q^2+Q_s^2}\quad .
\end{eqnarray*}
 Taking the derivative of $R$ and $R^D$ with respect to $Q^2$ and equating
it
to zero, we find that the ratio $R(Q^2)$ has no
maximum, while the maximum of the ratio $R^D(Q^2)$ is at
$Q^2_{max}=\sqrt{m_f^2 Q_s^2}$. There  is a  maximum of $R(Q^2)$
which  appears
when we take the
  logarithmic contributions into account ($R$
decreases logarithmically with $Q^2$) and turns out to be
proportional to $Q_s^2$. Our simple arguments show
that
\begin{enumerate}
\item\quad The maximum of $R$ is much shallower than that of
$R^D$\quad.
\item\quad The maximum of $R^D$ is proportional to $\sqrt{Q_s^2}$\quad.
\item\quad Existence of the non-perturbative scale $m_f^2$ is
crucial for appearance of maxima.
\end{enumerate}
 The maxima that we have just discussed in general occur due
to the well-known limiting behaviour of the longitudinal and
transverse cross sections at small and large $Q^2$. What we showed
is that at small enough $x_B$ those maxima are situated at a
scale, which can be expressed unambiguously in terms of the saturation
scale.

To check these conclusions, we plot the $Q_{max}^2$ and $Q_s^2$
versus $\log 1/x_B$ in Fig.~\ref{fig4}(a). Recall, that we obtained
formulae \eq{LONG}
and \eq{TRANS} by assuming that all dipoles in the virtual photon wave
function are of the size $4/Q^2$. Thus, to get the correct (up to
logarithmic contributions) numerical value of the saturation scale at
fixed $x_B$ we evaluated
the gluon distribution   at sufficiently large value of $Q^2$ (in fact
we took $Q^2=100$ GeV$^2$) where to a good approximation it is a constant.

 Note, that in making the simple estimations above we did not rely
on a   particular model. So, we also check our conclusions using
the
Golec-Biernat--Wusthoff model\cite{WUST} which is similar to ours,
but for which  we know exactly what  the
saturation scale is.  Using their $\hat\sigma$ we calculated the
ratios $R$ and $R^D$ and found the maxima.  The result is shown in
Fig.~\ref{fig4}(b). Surprisingly, both results agree quite well
with our rough estimations\footnote{Recently it was shown in
Ref.~\cite{KSH} that the experimental
data at small $x_B$  indeed exhibit scaling,  the total cross
section depends on the unique scaling variable $\tau=Q^2/Q_s^2$.}
(estimations of Fig.~\ref{fig4}(a) are valid at not too large $\log Q^2$).
Moreover, the maximum of $R$ is proportional  to $Q_s^2$ in
contrast to the maximum of $R^D$. This result can also be obtained from
our simple estimation  by including logarithmic corrections
in $\sigma_T$.

\section{Numerical estimations}

We now present the results of the numerical calculations. We
performed the calculation using GRV'94
parameterization\cite{GRV94,GRV98} for the gluon structure
function $x_BG(x_B,Q^2)$ for two reasons. First, we hope that the
non-linear corrections to the DGLAP evolution equation are  not
included in it. Second, it enables us  to perform calculations at values
of
$Q^2$ as small as $0.4$ GeV$^2$.

We noted in Ref.~\cite{WE} that the serious  drawback of GRV'94 is
the existence of a domain in
$Q^2$ where it gives the anomalous dimension of $x_BG(x_B,Q^2)$
larger than 1. To overcome this we use the following amended gluon
structure function
\begin{eqnarray}
x_BG(x_B,Q^2)&=&x_BG^{GRV}(x_B,Q^2)\:\theta(Q^2-\tilde Q^2(x_B))
\nonumber\\
&&+
\frac{Q^2}{\tilde Q^2(x_B)}\: x_BG^{GRV}(x_B,\tilde Q^2(x_B))
\:\theta(\tilde Q^2(x_B)-Q^2)\label{TILDQ}
\end{eqnarray}
where $\theta$ is a step function, and $\tilde Q^2(x_B)$ is a
scale at which the anomalous dimension equals 1 (it is a
saturation scale within our model accuracy). \eq{TILDQ} 
can be used as a
simplified parameterization of the behavior of the gluon structure
function in the saturation region. Our results turn out not to be
sensitive to the
 precise form of the behaviour of $xG(x_B,Q^2)$ in the vicinity of the
 saturation scale.

In order to compare our calculations with the
experimental data one has to multiply the theoretical formulae for the
cross sections (structure functions) by factor $\approx 0.5$. This factor
stems  from the next-to-leading order corrections which  can be
taken  into account by modeling the anomalous dimension as described in
Ref.~\cite{BRAZ}.

\begin{figure}
\begin{flushleft}
\begin{tabular}{ccc}
\multicolumn{3}{c}{\rule[-3mm]{0mm}{4mm} $ F_L(Q^2)$}\\
$x_B=10^{-3}$&$x_B=5\cdot 10^{-4}$& $x_B=10^{-5}$\\[-10mm]
\epsfig{file=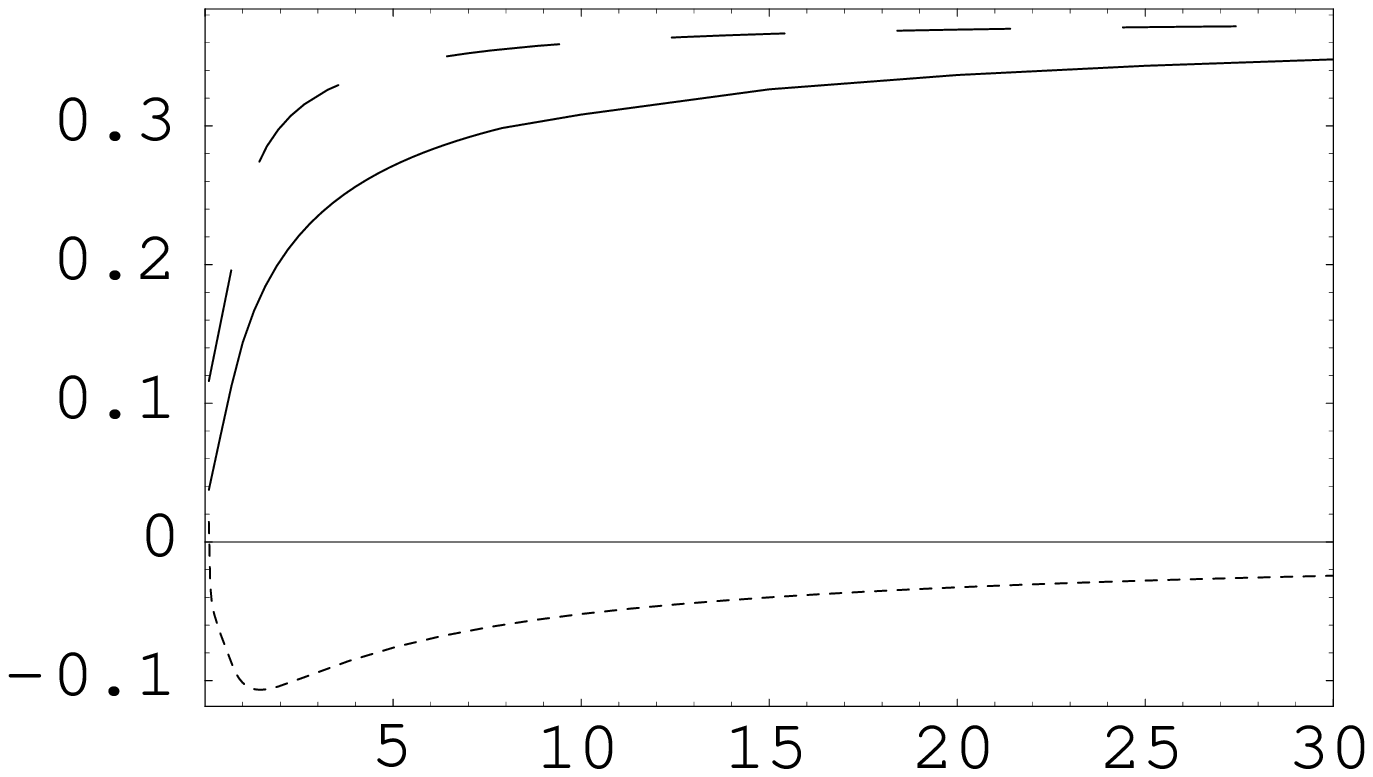,width=50mm,height=50mm}
& \epsfig{file=Pl_3.eps,width=50mm,height=50mm}&
\epsfig{file=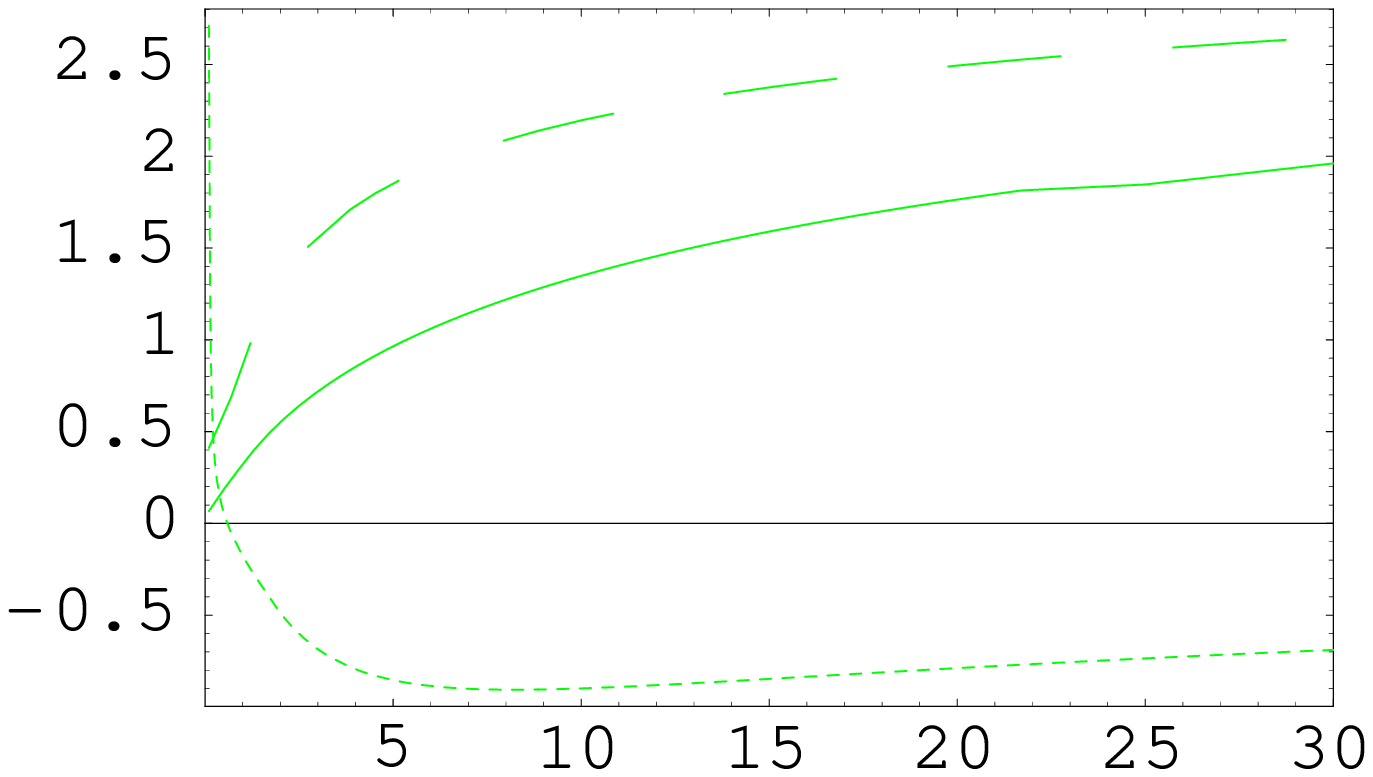,width=50mm,height=50mm}\\

\multicolumn{3}{c}{\rule[-3mm]{0mm}{4mm} $F_T(Q^2)$}\\
$x_B=10^{-3}$&$x_B=5\cdot 10^{-4}$& $x_B=10^{-5}$\\[-10mm]
\epsfig{file=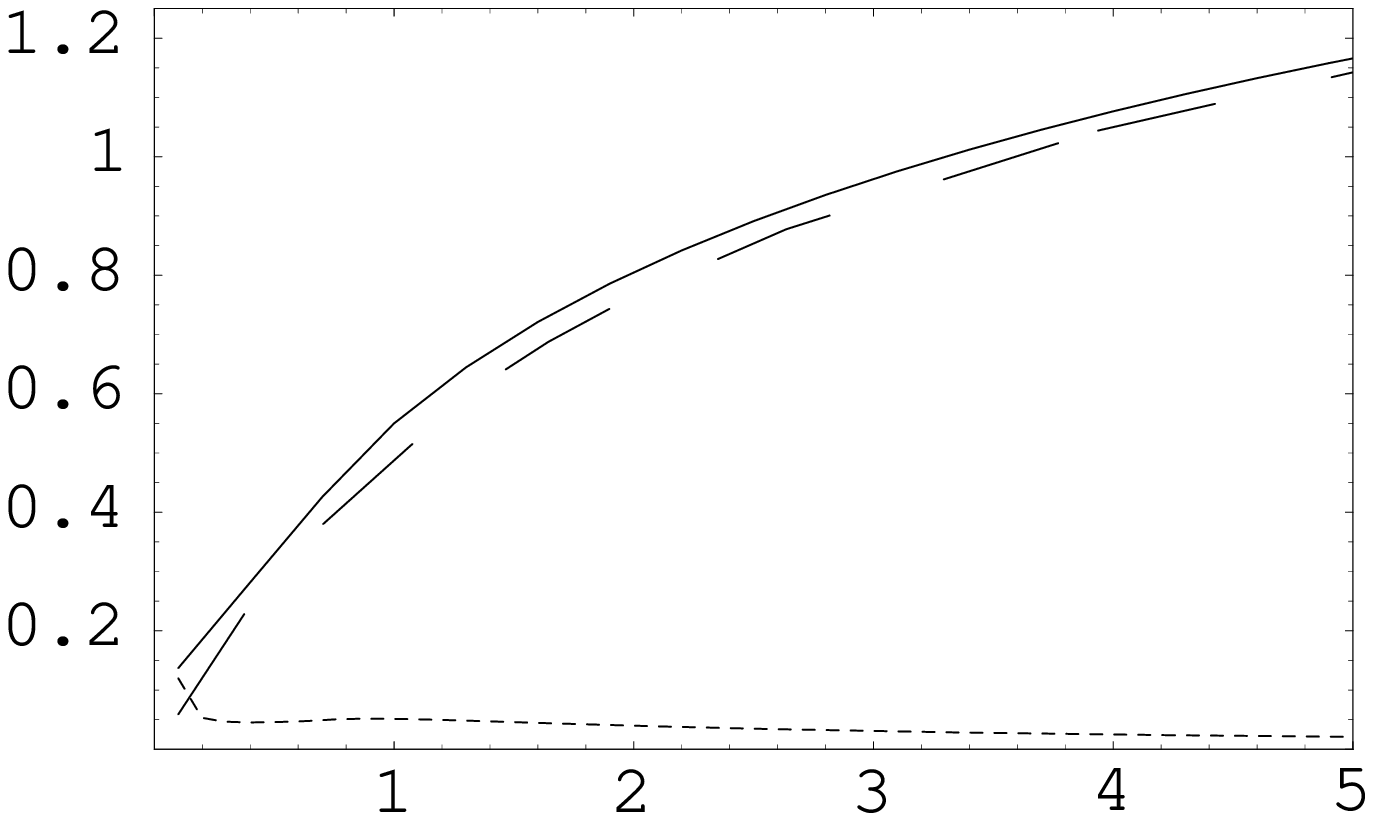,width=50mm,height=50mm} &
\epsfig{file=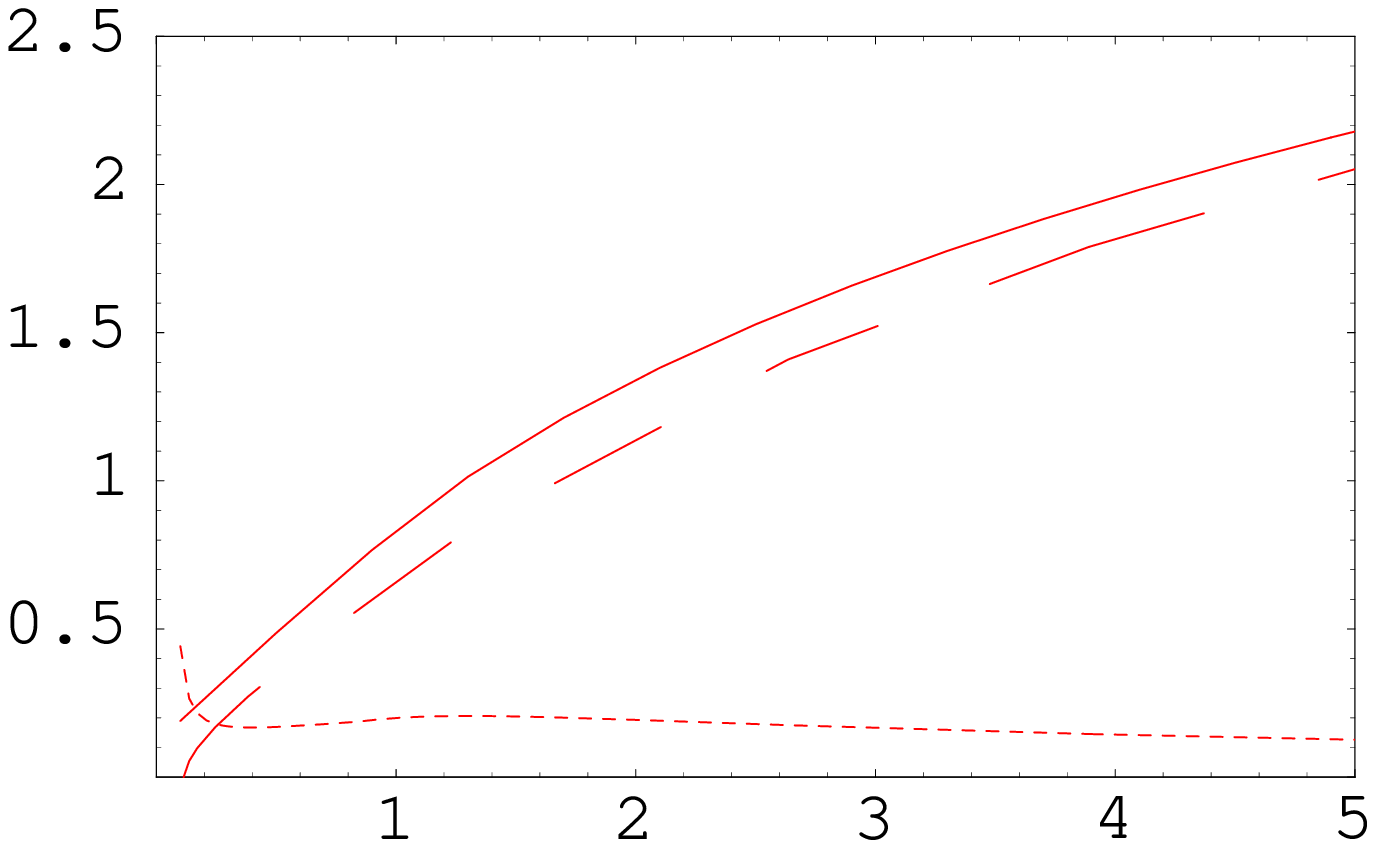,width=50mm,height=50mm} &
\epsfig{file=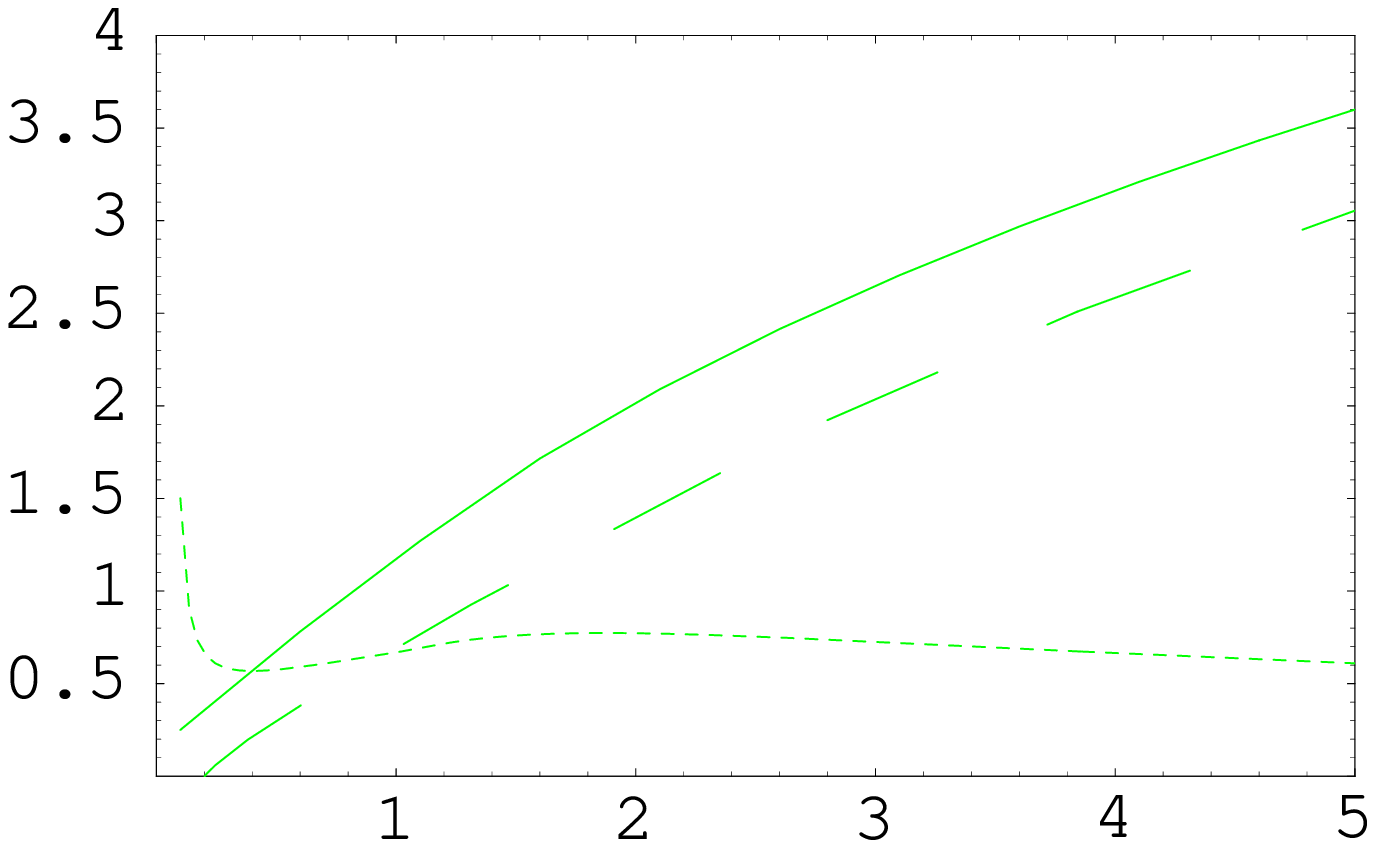,width=50mm,height=50mm}\\

\multicolumn{3}{c}{\rule[-3mm]{0mm}{4mm} $F_L^D(Q^2)$}\\
$x_B=10^{-3}$&$x_B=5\cdot 10^{-4}$& $x_B=10^{-5}$\\[-10mm]
\epsfig{file=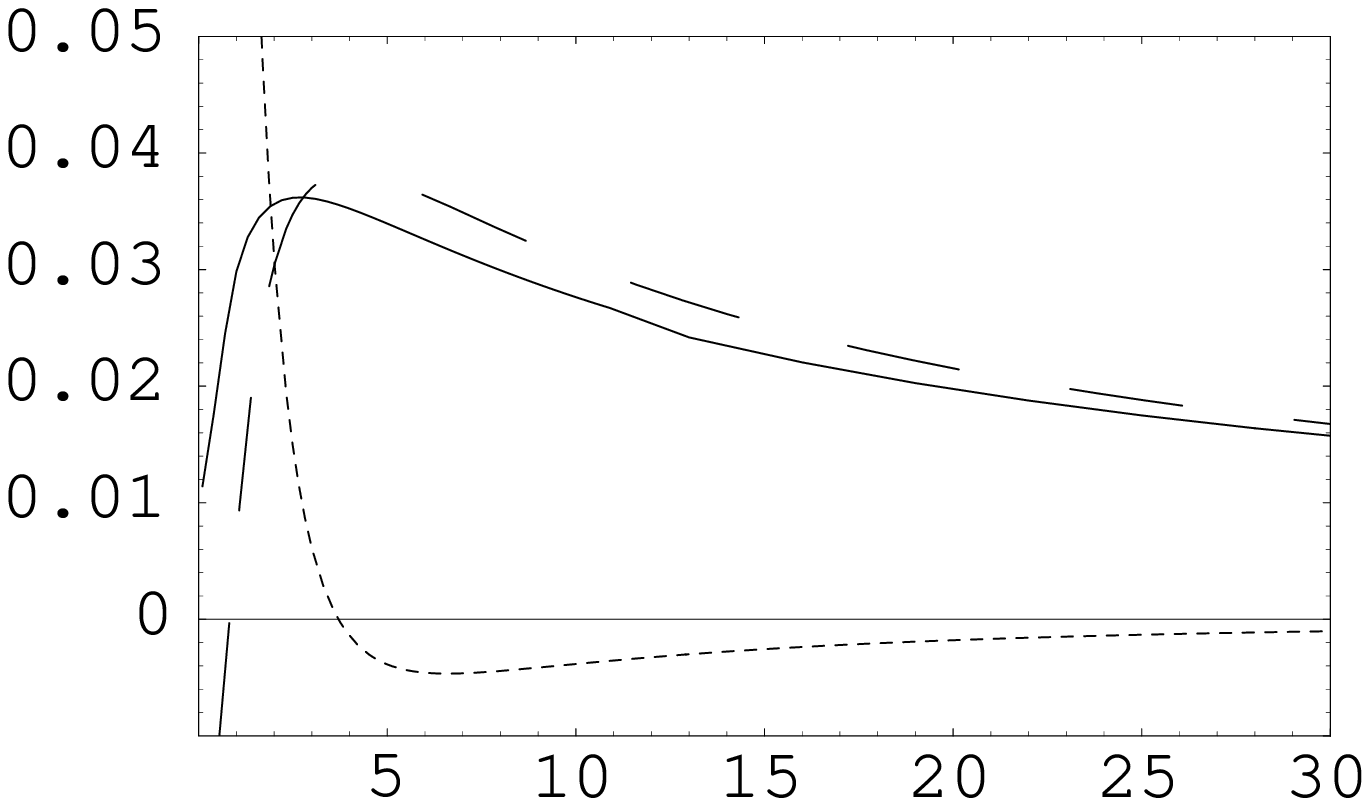,width=50mm,height=50mm} &
\epsfig{file=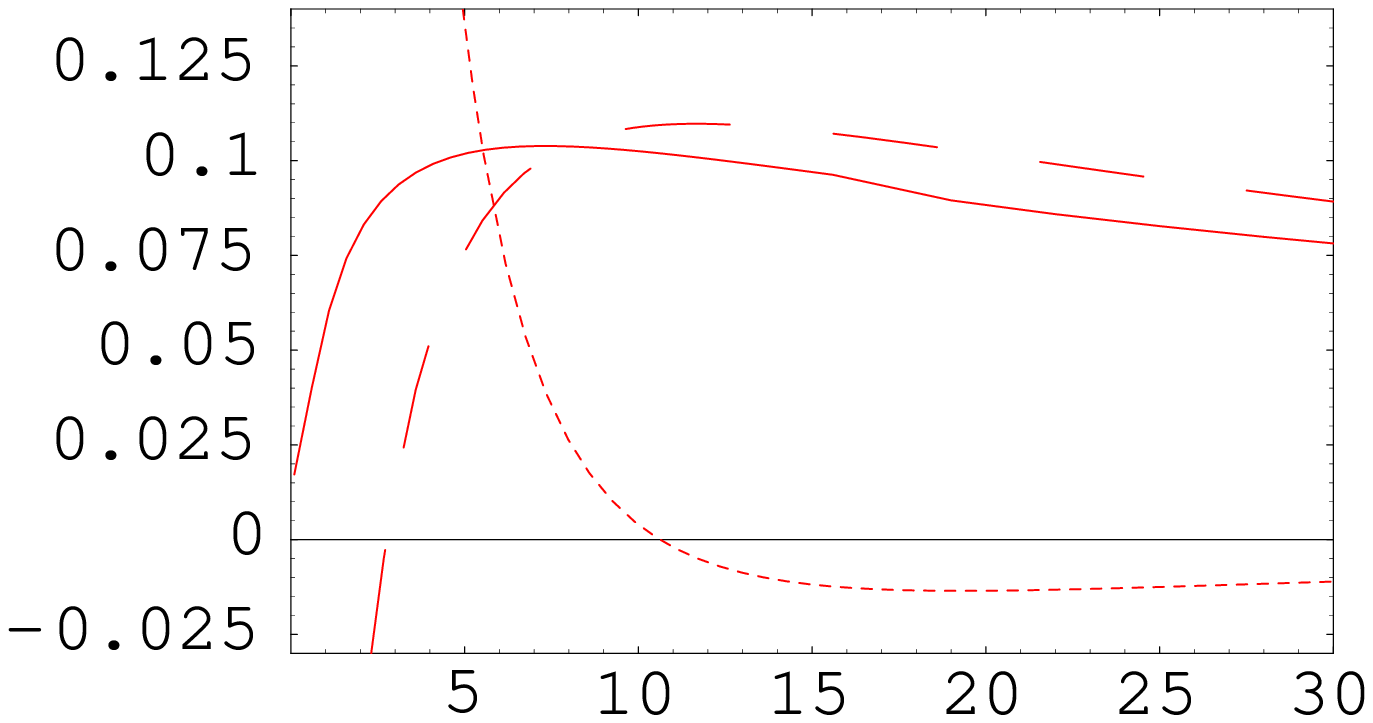,width=50mm,height=50mm} &
\epsfig{file=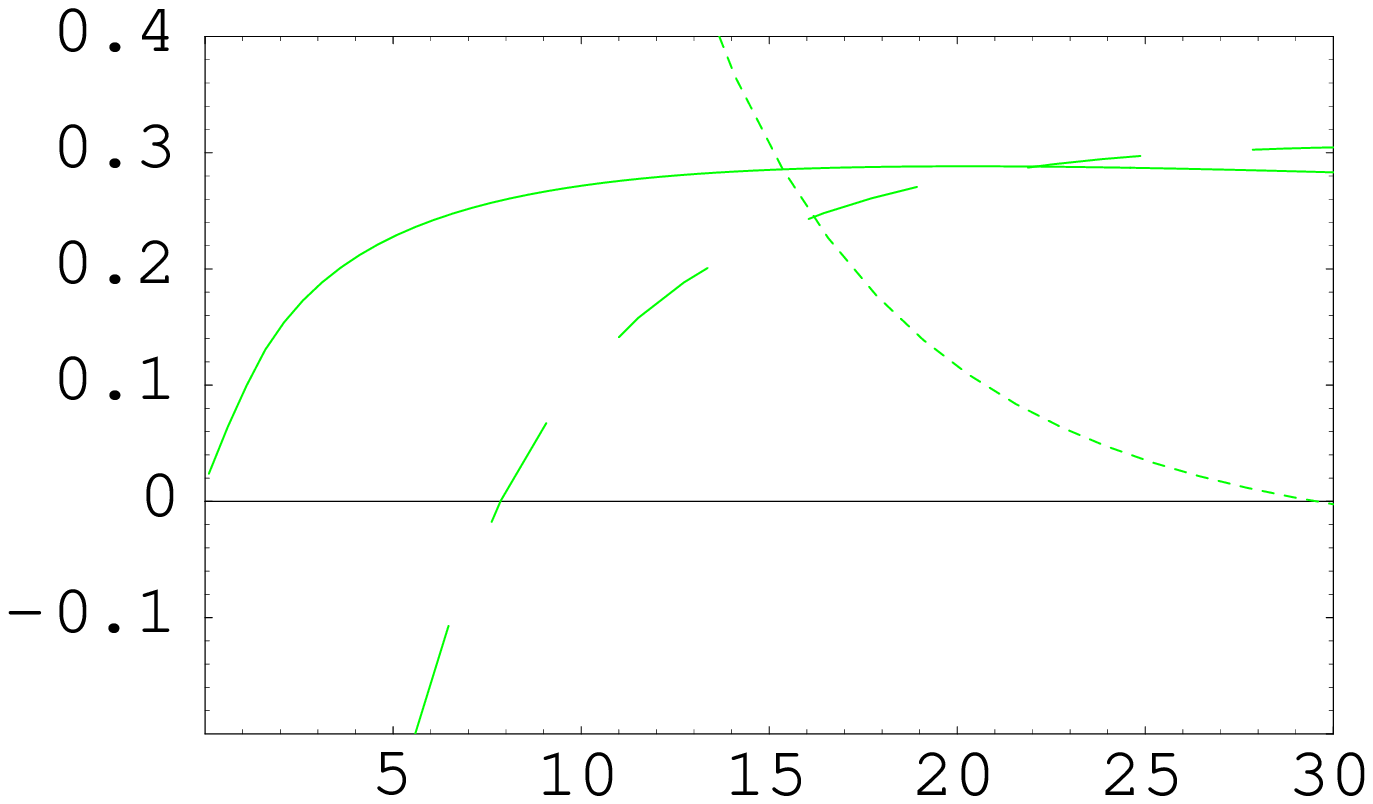,width=50mm,height=50mm}\\

\multicolumn{3}{c}{\rule[-3mm]{0mm}{4mm} $F_T^D(Q^2)$}\\
$x_B=10^{-3}$&$x_B=5\cdot 10^{-4}$& $x_B=10^{-5}$\\[-10mm]
\epsfig{file=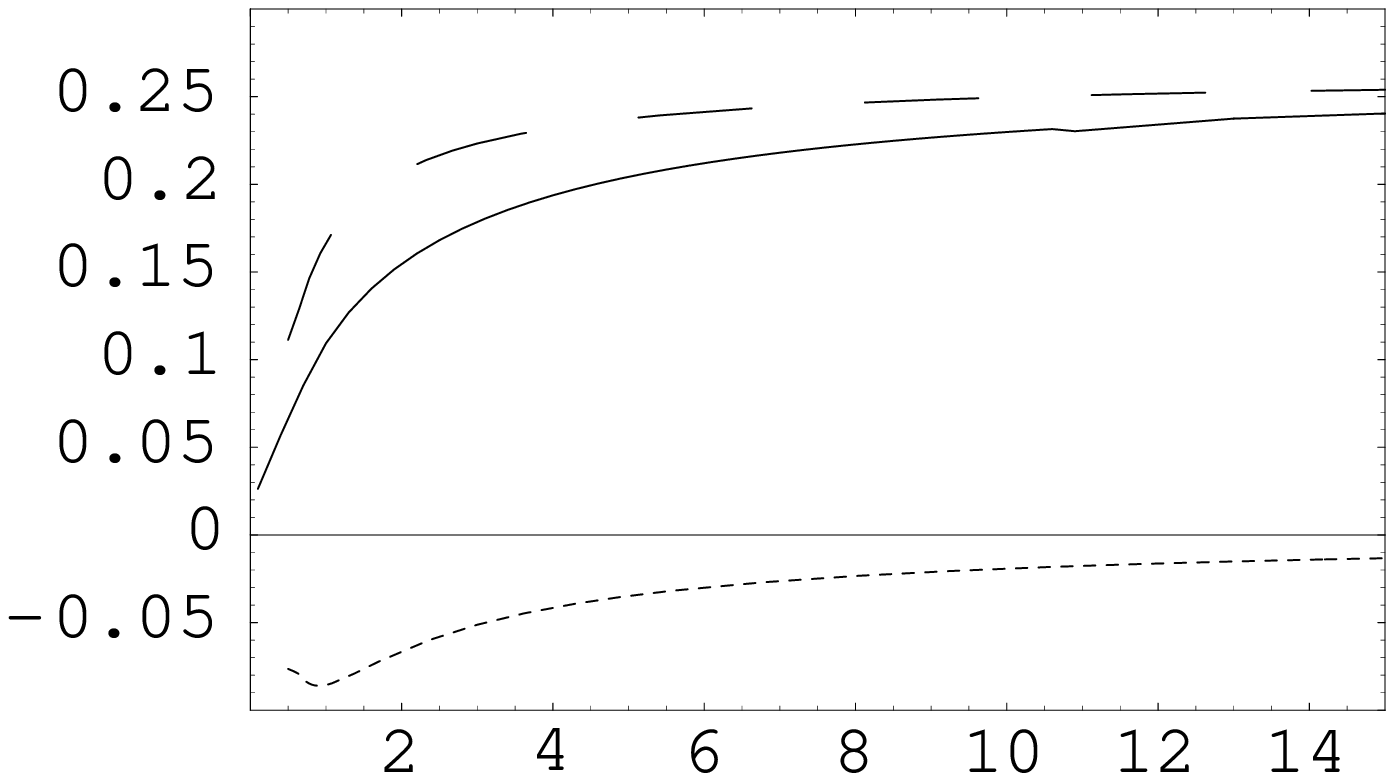,width=50mm,height=50mm} &
\epsfig{file=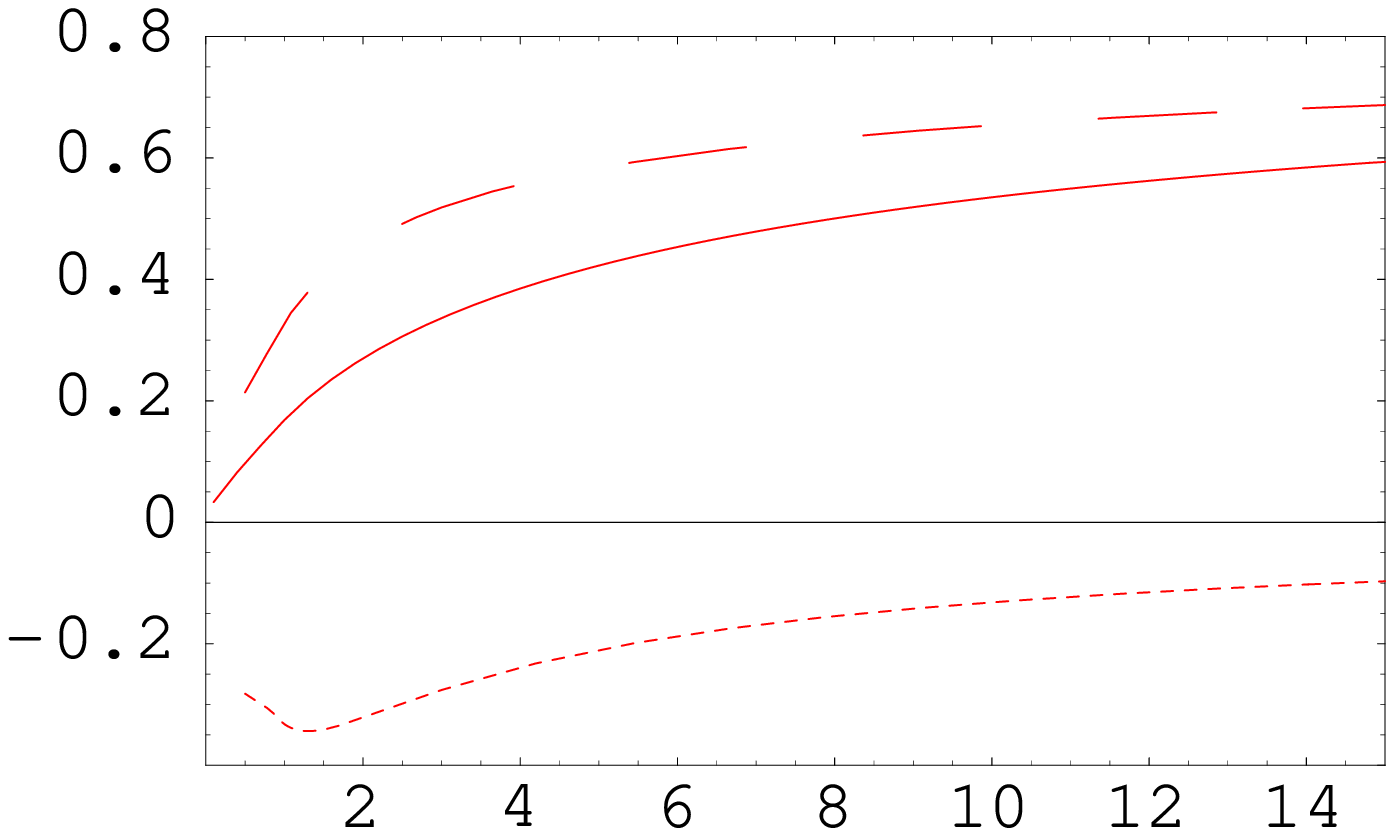,width=50mm,height=50mm} &
\epsfig{file=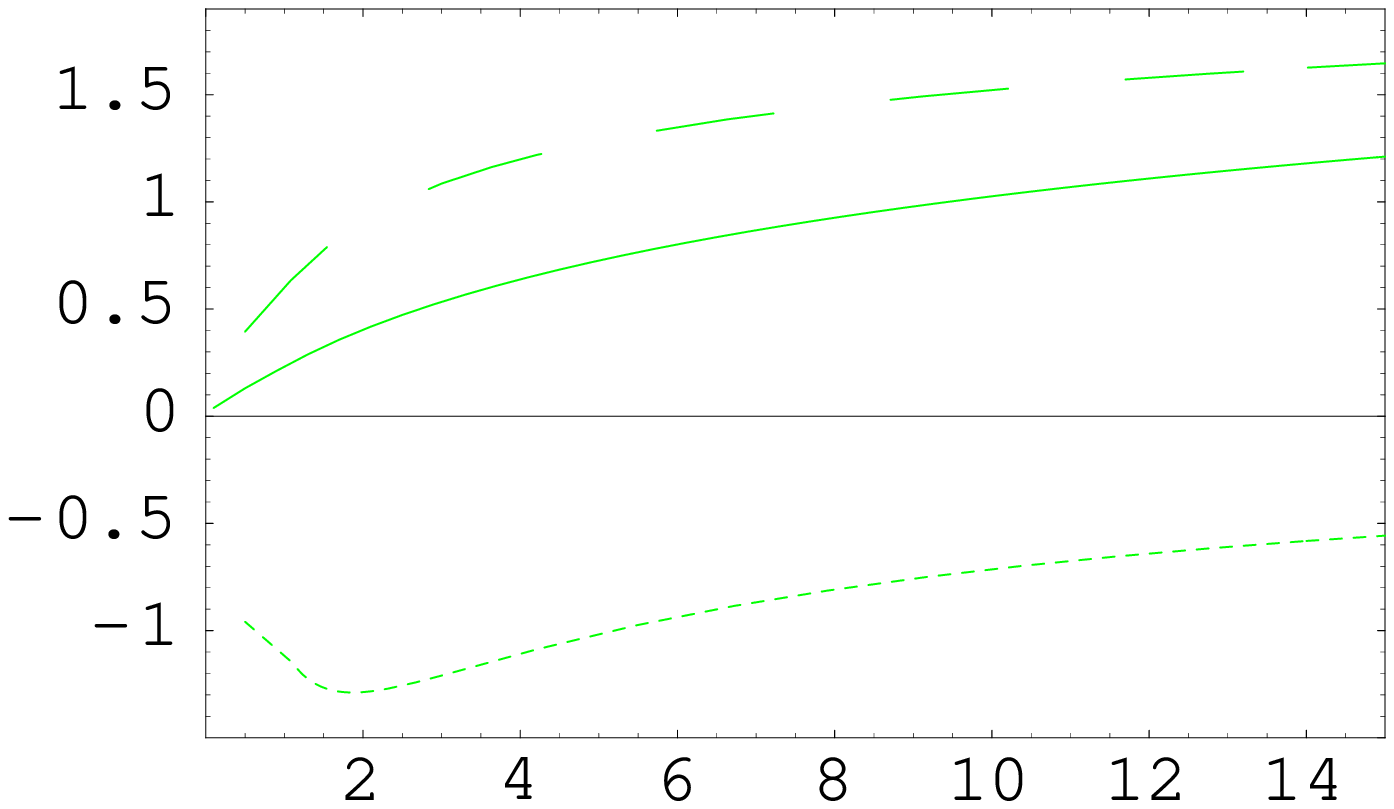,width=50mm,height=50mm}
\end{tabular}
\end{flushleft}
\vspace{-1cm}
\caption{\footnotesize {\it Different twist contributions
to the various structure functions for DIS on the proton:
leading twist (at high $Q^2$) -- dashed line,
next-to-leading -- dotted one, exact structure function --
solid curve.}}
\label{fig2.30}
\end{figure}

To estimate contributions of different twists to various structure
functions defined by
 $$
F(x_B,Q^2)=\frac{Q^2}{4\alpha_{em}\pi^2}\sigma(\gamma^*p) \quad,
 $$
 we performed calculations using \eq{LONG}, \eq{TRANS} and
\eq{DIFFRACT}. Note, that \eq{LONG} and \eq{TRANS} are valid only
in the  massless limit $m_f^2=0$. As was explained above, this is
a reasonable assumption as long as we are not concerned with 
the behavior at small $Q^2$.
 All calculations are made for the THERA kinematical
region. The results of calculations for different twist
contributions are shown in Fig.~\ref{fig2.30}.
Similarly to the nuclear target case the following
important effects are  seen, when we compare contributions
of the first two non-vanishing twists:
\begin{enumerate}
\item\quad In all structure functions except $F_L$ there  is a scale $\hat
Q^2$  for the particular structure function where these two twist
corrections are
equal. $\hat Q^2$ is largest for $F_L^D$;

\item\quad $\hat Q^2$ grows as $x_B$ decreases;

\item\quad Twist-4 corrections in $F_L$ cancels numerically with twist-4
in\footnote{This effect was   also noted in Ref.~\cite{BARTELS} where
another, though similar, model\cite{WUST} for SC was used.} $F_T$ in
the region where their separate contributions are quite large.
This is a reason to measure the polarized structure functions
instead of the total one.
\end{enumerate}

 In analogy with the scale $Q^2_{max}$ the scale at which OPE
breaks down is expected to be a certain function of the saturation
scale,  this function  differs for the various structure
functions. The question of what  the analytic expression of this
function is warrants further study. We, however, can assert that the
scale at which the
OPE breaks down  is not smaller than $\hat Q^2$. The latter can
 easily be estimated by equating the analytic expressions for twist-2
and twist-4 corrections
obtained in Ref.~\cite{WE}. The result is $\hat
Q^2$
$=f\cdot Q^2_{s}$ up to logarithmic corrections, where the
numerical factor $f$ is of order unity (e.g. for
$\sigma_T$, $f=0.79$).

\begin{figure}
\begin{flushleft}
\begin{tabular}{cc}
\epsfig{file=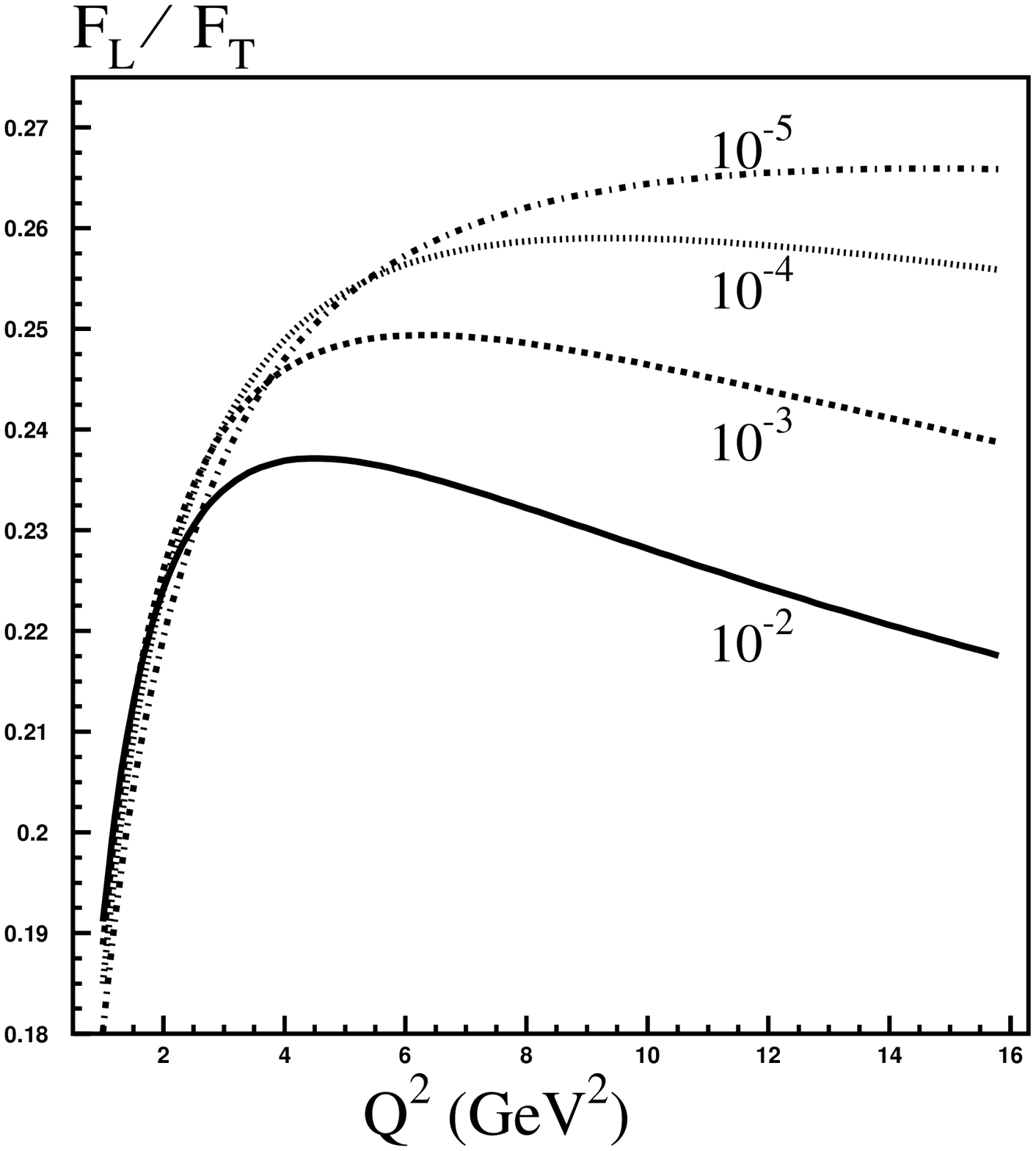,width=80mm,height=80mm}
&
\epsfig{file=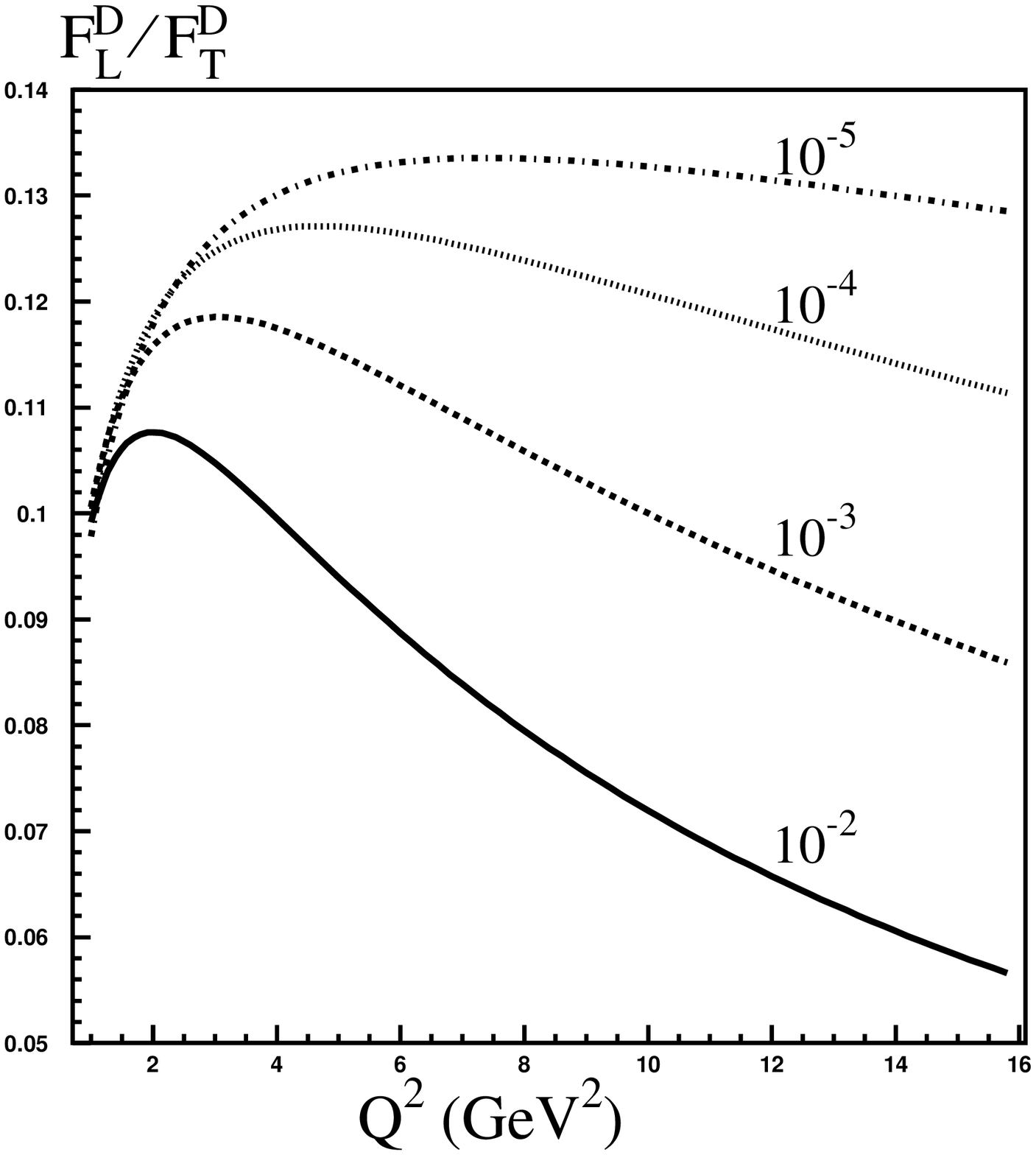,width=80mm,height=80mm}\\
$(a) $& $(b)$
\end{tabular}
\end{flushleft}
\vspace{0.17cm}
\caption{\footnotesize {\it $(a)$ Ratio $F_L/F_T$ versus $Q^2$ for
different $x_B$, $(b)$ the same for $F_L^D/F_T^D$ }}
\label{fig3}
\end{figure}

In Figs.~\ref{fig3} we plot the ratios
$F_L/F_T\,(x_B,Q^2)$ and $F_L^D/F_T^D\,(x_B,Q^2)$  versus $Q^2$. The
calculations were performed without neglecting the quark mass $m_f$.
As is expected on theoretical grounds the ratio has
maximum which increases as $x_B$ decreases\cite{GLR}. In the kinematical
region of
THERA it is expected, that $Q_s^2\sim 1/x_B$\cite{LT}.
The maxima in Fig.~\ref{fig3} occur at $Q^2_{max}\gg m_f^2$ which
implies that we are sufficiently far from the confinement
region. The maximum of the ratio of the diffractive structure functions
$R^D$
is  more obvious, as we saw in sec.~2.

We show in Figs.~\ref{fig4}(a) the  behavior of $Q^2_{max}(x_B)$
as a function of $x_B$.  The increase is still smaller than predicted
theoretically. Our conclusion is that the kinematical
region of THERA is intermediate between the linear and saturation regimes.

\section{Conclusions}
\begin{figure}
\begin{flushleft}
\begin{tabular}{cc}
\epsfig{file=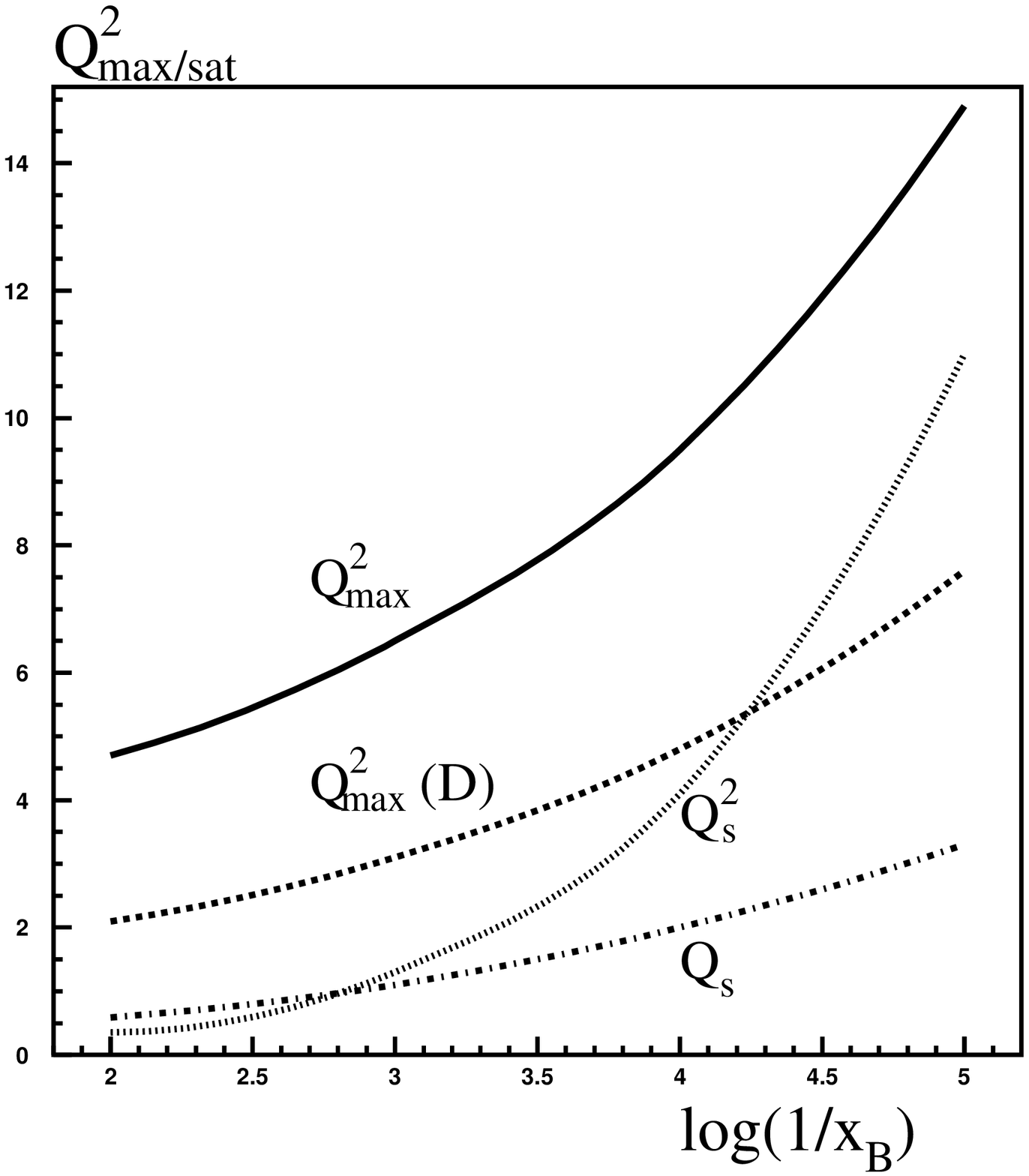,width=80mm,height=80mm}
&
\epsfig{file=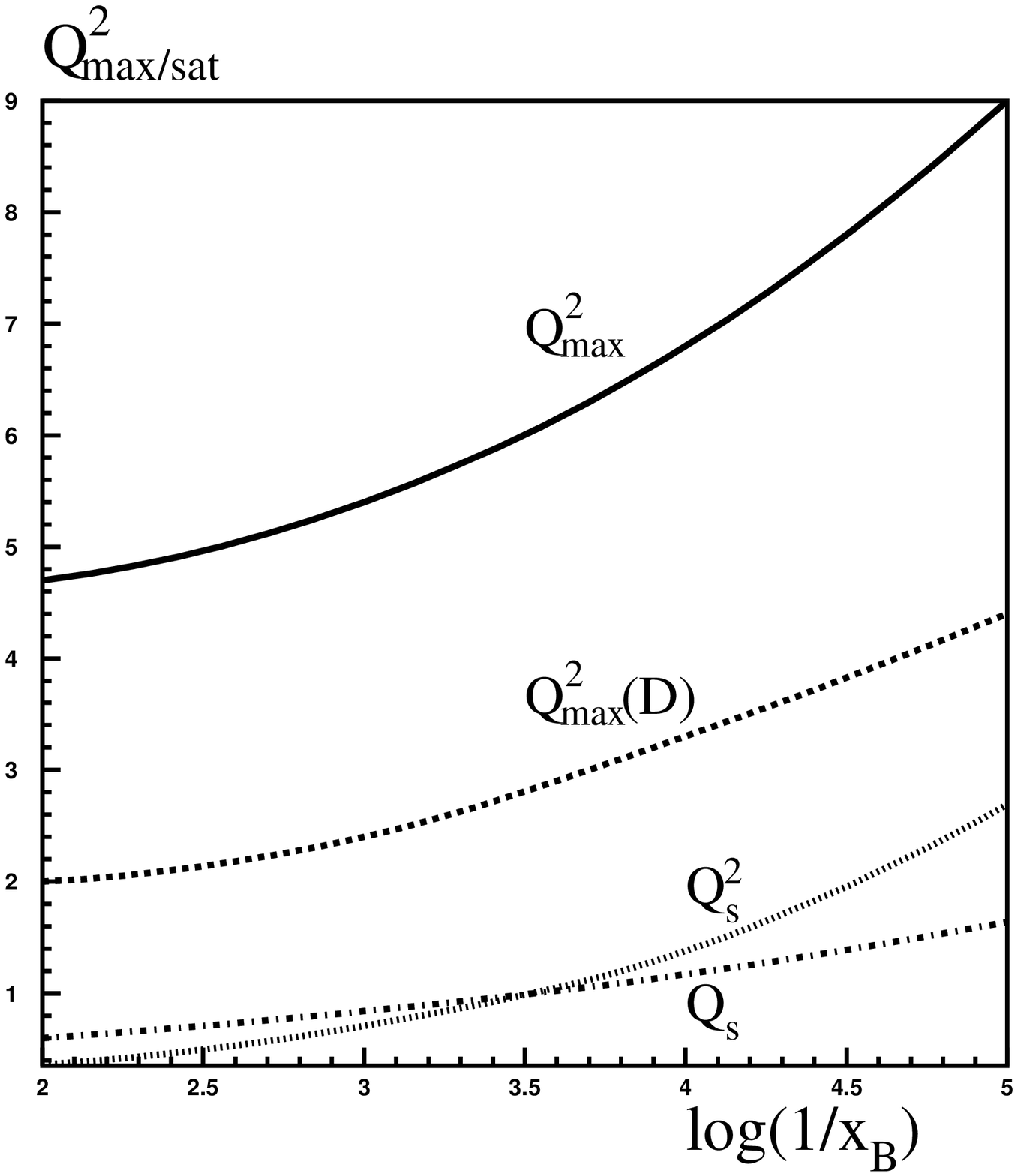,width=80mm,height=80mm}\\
$(a) $& $(b)$
\end{tabular}
\end{flushleft}
\vspace{0.17cm}
\caption{\footnotesize {\it Scaling of the maxima of ratios $F_L/F_T$
($Q^2_{max}$) and  $F_L^D/F_T^D$ ($Q^2_{max}(D)$) with
$\log_{10}(1/x_B)$
in $(a)$ the eikonal model, $(b)$ the GW model. }}
\label{fig4}
\end{figure}

The aim of our work is to understand to what extent the high parton
density regime of QCD could be detected in DIS experiments at
HERA, there the density of partons could be sufficiently large  to
make SC
significant.
Using the eikonal model we observed two manifestations of the saturation
scale:
\begin{enumerate}
\item\quad The position of the maxima $Q_{max}^2$ in ratios $F_L/F_T$
and  $F_L^D/F_T^D$ is intimately related to the saturation scale
$Q_s^2(x_B)$.  $Q_{max}^2$ shows $x_B$ dependence typical for the
saturation scale (see Fig.~\ref{fig4}). On the contrary, the
ratios of structure functions satisfying DGLAP equation have no maxima,
as is readily seen from the leading twist formulae given in
Ref.~\cite{WE}.

\item\quad The maximum of the ratio $F_L^D/F_T^D$ is much more apparent
than that of $F_L/F_T$, since the former occurs due to leading twist  
behaviour of the structure functions, whereas the later is due to
logarithmic
corrections to this behaviour.

\item\quad The results of our calculations show that there exists a
scale at which  the
twist expansion breaks down since all twist corrections become of the same
order.
We can use the DGLAP evolution equations only for $Q^2$ larger than this
scale. The energy ($x_B$) dependence of this scale suggests
that the experimental data for deep inelastic scattering on a  proton will
help us  separate leading and higher twist contributions. In the case
of nucleon deep inelastic scattering such a separation appears to be a
rather difficult task and  has not yet been performed.
\end{enumerate}

\vskip0.3cm
{\large\bf Acknowledgments}
\vskip0.3cm

We thank Jochen Bartels,  Krzysztof Golec-Biernat, Dima Kharzeev and Yura
Kovchegov for  very fruitful discussions concerning saturation phenomena.

This paper in part was supported by  BSF grant  \# 9800276.
The work of L.McL.  was supported by the US Department of Energy
(Contract \# DE-AC02-98CH10886).
The research of E.L. and K.T. was supported in part by the Israel Science
Foundation, founded by the Israeli Academy of Science and Humanities.

E.G. and E.L. are very grateful to DESY Theory Group for hospitality
extended to them during their stay at DESY where this paper was completed.

\newpage
%
%

\end{document}